\documentstyle[hutp]{article}

\makeatletter \input art10.sty \makeatother
\def\starttext{\twocolumn}

\newcommand{\beq}{\begin{equation}}
\newcommand{\eeq}{\end{equation}}

\newcommand{\remove}[1]{}

\input prepictex.tex
\input pictex.tex
\input postpictex.tex

\newcommand{\ds}{\displaystyle}

\newcommand{\be}{\begin{equation}}
\newcommand{\ee}{\end{equation}}
\newcommand{\bp}{\begin{picture}}
\newcommand{\ep}{\end{picture}}

\def\lte{\mathrel{\displaystyle\mathop{\kern 0pt <}_{\raise .3ex
\hbox{$\sim$}}}}
\def\gte{\mathrel{\displaystyle\mathop{\kern 0pt >}_{\raise .3ex
\hbox{$\sim$}}}}

\newcommand{\ba}[1]{\begin{array}{#1}\ds }

\newcommand{\ea}{\end{array}}

\def\lta{\mathrel{\displaystyle\mathop{\kern 0pt <}_{\raise .3ex
\hbox{$\sim$}}}}
\def\gta{\mathrel{\displaystyle\mathop{\kern 0pt >}_{\raise .3ex
\hbox{$\sim$}}}}

\newsavebox{\phru}
\savebox{\phru}{\beginpicture
\setcoordinatesystem units <\unitlength,\unitlength>
\setquadratic
\plot
0 0
2.5 3
5 0
7.5 -3
10 0
/
\endpicture}

\newsavebox{\phrd}
\savebox{\phrd}{\beginpicture
\setcoordinatesystem units <\unitlength,\unitlength>
\setquadratic
\plot
0 0
2.5 -3
5 0
7.5 3
10 0
/
\endpicture}

\newsavebox{\phdr}
\savebox{\phdr}{\beginpicture
\setcoordinatesystem units <\unitlength,\unitlength>
\setquadratic
\plot
0 0
3 -2.5
0 -5
-3 -7.5
0 -10
/
\endpicture}

\newsavebox{\phdl}
\savebox{\phdl}{\beginpicture
\setcoordinatesystem units <\unitlength,\unitlength>
\setquadratic
\plot
0 0
-3 -2.5
0 -5
3 -7.5
0 -10
/
\endpicture}

\newdimen\pmboffset
\def\oldpmb#1{\setbox0=\hbox{#1}%
 \copy0\kern-\wd0
 \kern\pmboffset\raise 1.732\pmboffset\copy0\kern-\wd0
 \kern\pmboffset\box0}

\begin{document}

\title{Flavor Changing Neutral Currents in a Realistic Composite
Technicolor Model}

\author{
        Christopher D. Carone and Rowan T. Hamilton \thanks {
        carone@huhepl, hamilton@huhepl
        }  \\
        Lyman Laboratory of Physics \\
        Harvard University \\
        Cambridge, MA 02138
}
\date{\today}

\renewcommand{\preprintno}{HUTP-A059}

\begin{titlepage}

\maketitle

\def\thepage {}        

\begin{abstract}
We consider the phenomenology of a composite technicolor model
proposed recently by Georgi.  Composite technicolor interactions produce
four-quark operators in the low energy theory that contribute to flavor
changing neutral current processes.  While we expect operators of this type
to be induced at the compositeness scale by the flavor-symmetry breaking
effects of the preon mass matrices, the Georgi model also includes operators
from higher scales that are not GIM-suppressed.  Since these operators
are potentially large, we study their impact on flavor changing neutral
currents and CP violation in the neutral $B$, $D$, and $K$ meson systems.
\end{abstract}

\end{titlepage}

\starttext 
\pagestyle{columns} 
\pagenumbering {arabic} 

\section {Introduction} \label {sec:intro}

Minimal technicolor \cite{mtech} models present an elegant mechanism
for electroweak symmetry breaking, but they do not explain how
the fermions acquire their masses.  Extended technicolor (ETC)
models \cite{etech} provide a mechanism for fermion mass generation
by means of extended gauge interactions that couple the quarks and
leptons to technifermion condensates.  Since these interactions mediate
flavor changing neutral currents (FCNC), however, it is difficult
to construct a realistic ETC model that can generate a large top quark
mass without violating the FCNC bounds.  Composite technicolor standard
models (CTSM) \cite{ctech} attempt to overcome this problem by introducing
three different ETC groups, so that it is possible to maintain a GIM
mechanism in the CTSM sector of the theory.  The flavor symmetry breaking
effects of the preon mass matrices in CTSM models are responsible for
generating the fermion masses, while producing FCNC effects at acceptable,
and often interesting, levels.

In ref. \cite{moose}, a composite technicolor model was presented
that approximately reproduces the correct phenomenology below
$1.5$ TeV.  In this model, the operators that contribute to flavor
changing neutral current processes come not only
from the symmetry breaking effects of $M_{\rm preon}$ at the
compositeness scale, $f_2$, but also from operators,
produced at a higher scale $f_1$, that are not GIM suppressed.
These new operators are of the form
\beq
\frac{1}{f_1^2}(\overline{u}_L \gamma^\mu V S^\dagger u_L)
(\overline{u}_L \gamma^\mu S V^\dagger u_L)
\label{eq:opup}
\eeq
for the charge $2/3$ quarks, and
\beq
\frac{1}{f_1^2}(\overline{d}_L \gamma^\mu S^\dagger V d_L)
(\overline{d}_L \gamma^\mu V^\dagger S d_L)
\label{eq:opdown}
\eeq
for the charge $-1/3$ quarks, where $V$ is the CKM matrix, and $S$
is a matrix that is determined by the detailed dynamics of the model.
If we write $V$ in the parameterization of ref. \cite{moose}
\beq
V = \left( \begin{array}{cc}
e^{i(\beta-\alpha)}(1-(1-c_\phi)uu^\dagger) \Sigma & -s_\phi u \\
e^{i\beta} s_\phi u^\dagger \Sigma & e^{i\alpha} c_\phi
\end{array}
\right)
\label{eq:CKM}
\eeq
where $\Sigma$ is a unitary, unimodular $2\times 2$ matrix, and
$u$ is a two component complex vector, then the matrix $S$ is identical
to (\ref{eq:CKM}) with $\phi$ replaced by an angle
$\phi'$, that is determined by the model's vacuum alignment below the scale
$f_2$.   We refer the reader to the original literature for
technical details.  The operators (\ref{eq:opup}) and (\ref{eq:opdown})
alter the results of the standard model box diagram calculation for the
$\Delta F =2$ amplitudes, where $F=B$, $C$, or $S$.  In this letter, we
consider how the totality of CTSM interactions in the model of
ref. \cite{moose} alters the standard model predictions for flavor
mixing and CP violation in the neutral $B$, $D$, and $K$ meson systems.
Our analysis establishes a definite pattern of deviations from the
standard model predictions that may be tested by measurements at future
B factories, and by a more precise measurement of $\epsilon '/\epsilon$.

\section{B-Meson Physics} \label {sec:bphys}

The box-diagram contribution to the operator $(\overline{b}_L \gamma^\mu
d_L)^2$ in the standard model is given by \cite{inami}
\beq
\frac{G_F^2}{16\pi^2}\left[(\xi^B_t)^2 m_t^2 + \xi^B_t \xi^B_c m_c^2
\ln \left(\frac{m_c}{m_t}\right)^2 + (\xi^B_c)^2 m_c^2 \right]
\label{eq:bdbox}
\eeq
where
\beq
\xi^B_q \equiv V_{qd} V_{qb}^*
\eeq
For a top quark mass on the order of 100 GeV, the three terms in
(\ref{eq:bdbox}) have magnitudes in the ratio $10^3$::$10$::1, so we
may safely neglect the last two terms in comparison to the first,
\beq
{\cal O} ^{\Delta B =2}_{{\rm S.M.}} \approx
\frac{G_F^2}{16\pi^2} (\xi^B_t)^2 m_t^2 (\overline{b}_L \gamma^\mu
d_L)^2
\label{eq:smbd}
\eeq
A generic CTSM model contributes to the $\Delta B = 2$ operator by
virtue of the flavor-symmetry breaking effects of the preon mass
matrices \cite{ctech,sekhar}.  Treating the the preon mass matrices as
spurions, we can write down the following operator, consistent with the
flavor symmetries of the model
\beq
\frac{1}{(4\pi)^4 f_2^6}(\overline{\psi}_L \gamma^\mu M_U M_U^\dagger
\psi_L)(\overline{\psi}_L \gamma^\mu M_U M_U^\dagger
\psi_L)
\label{eq:mumu}
\eeq
where $M_U$ is the preon mass matrix that is proportional to
the mass matrix of the charge $2/3$ quarks. The order of magnitude of
the coefficient of the operator has been determined by naive
dimensional analysis \cite{NDA}, but it's sign is unknown.  From
(\ref{eq:mumu}) we then find \cite{sekhar}
\beq
{\cal O}^{\Delta B = 2}_{1} \approx \frac{G_F^3}{(4\pi)^4}(\xi^B_t)^2
m_t^4 \left(\frac{f_2}{v}\right)^6 (\overline{b}_L \gamma^\mu
d_L)^2
\label{eq:ctsmbd1}
\eeq
where $v$ is the electroweak symmetry breaking scale, and where, as in
(\ref{eq:smbd}), we have only retained the leading terms.  The new
operator (\ref{eq:opdown}) gives us an additional contribution
\beq
{\cal O}^{\Delta B =2} _{2} = -\frac{1}{f_1^2}\, e^{2 i \alpha}\,(w_1^*)^2
\,\sin ^2(\phi - \phi ')\,(\overline{b}_L \gamma^\mu
d_L)^2
\label{eq:ctsmbd2}
\eeq
where, in the notation of (\ref{eq:CKM}), $w$ is a complex column vector
defined by
\beq
w \equiv \Sigma \, u
\label{eq:vdef}
\eeq
{}From now on we will adopt the parametrization of the CKM
matrix in which the largest phases appear in $V_{ub}$ and
$V_{td}$.  Then we may combine (\ref{eq:smbd}), (\ref{eq:ctsmbd1}),
and (\ref{eq:ctsmbd2}), to obtain
\beq
{\cal O}^{\Delta B =2} = \left[ \frac{G_F^2m_t^2}{16\pi^2} -
\frac{1}{f_1^2}\frac{\sin^2(\phi-\phi')}{\sin^2\phi} \pm
\frac{G_F^3 m_t^4}{(4\pi)^4}\left(\frac{f_2}{v}\right)^6 \right]
(V_{td})^2 (\overline{b}_L \gamma^\mu
d_L)^2
\label{eq:bdfull}
\eeq
A similar analysis gives us the operator responsible for the $B^0_S$-$
\overline{B}^0_S$ mass difference
\beq
{\cal O}^{\Delta B =2} = \left[ \frac{G_F^2m_t^2}{16\pi^2} -
\frac{1}{f_1^2}\frac{\sin^2(\phi-\phi')}{\sin^2\phi} \pm
\frac{G_F^3 m_t^4}{(4\pi)^4}\left(\frac{f_2}{v}\right)^6 \right]
(V_{ts})^2 (\overline{b}_L \gamma^\mu
s_L)^2
\label{eq:bsfull}
\eeq
The CTSM parameters in (\ref{eq:bdfull}) and (\ref{eq:bsfull}) were
estimated in ref. \cite{moose} by naive dimensional analysis.
Assuming the values, $f_1 \approx 130$ TeV, $f_2 \approx 1.4$ TeV,
$\phi ' \approx 1$, and $m_t \approx 100$ GeV, from ref. \cite{moose},
and $\phi \approx 0.04$ from the magnitude of $V_{tb}$, we may estimate
the three contributions to the $\Delta B = 2$ operators.  We find
\beq
{\cal O}^{\Delta B =2} = \left[
9\times 10^{-9} \, - \, 2\times 10^{-8} \,\pm \, 2\times 10^{-7}
\right] (\mbox{GeV}^{-2})(V_{tq})^2 (\overline{b}_L \gamma^\mu
q_L)^2
\label{eq:bdsizes}
\eeq
for $q=d$ or $s$.  Clearly the non-standard contributions
are likely to be as important as the contribution from
the standard model box diagram.  While it is tempting to make
a stronger statement based on the large size of the third term
in (\ref{eq:bdsizes}), our estimate of this contribution is by
far the most uncertain, given the the sensitive
dependence of (\ref{eq:ctsmbd1}) on the unknown scale $f_2$.
In any event, if we knew the top quark mass and had an accurate,
independent measure of $V_{td}$ and $V_{ts}$, then these results suggest
that the mass splittings in the $B^0$ and $B^0_S$ systems could differ
significantly from the standard model predictions, while remaining in
the expected ratio, $(V_{td}/V_{ts})^2 $.  Unfortunately, testing
this prediction would require independent measurement
of these CKM elements from the semileptonic decays of $T$ mesons, which
is unlikely anytime in the near future.

\begin{figure}
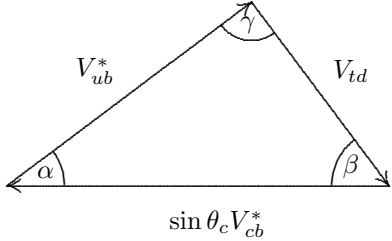

$$
\beginpicture
\setcoordinatesystem units < 0.02in, 0.02in > point at 0 0
\arrow <5pt> [0.3,1] from 100 0 to 0 0
\arrow <5pt> [0.3,1] from 0 0 to 64 48
\arrow <5pt> [0.3,1] from 64 48 to 100 0
\put{$\alpha$}  at 10 3.5
\put{$\beta$}  at 90  5
\put{$\gamma$}  at 63 42
\circulararc 36.86989 degrees from 15 0 center at 0 0
\circulararc 90.0 degrees from 56 42 center at 64 48
\circulararc -53.13011 degrees from 85 0 center at 100 0
\put{$\sin \theta_c V^*_{cb}$} at 55 -10
\put{$V_{td}$} at 90 30
\put{$V_{ub}^*$} at 23 30
\endpicture
$$
\label{fig:untri1}
\caption{Unitarity triangle in the standard model.}
\end{figure}

A more realistic method of uncovering for the effects of (\ref{eq:bdfull})
and (\ref{eq:bsfull}) is by testing the standard model unitarity relation
\beq
V_{ud} V_{ub}^* + V_{cd} V_{cb}^* + V_{td} V_{tb}^* = 0
\label{eq:ut}
\eeq
which defines a triangle in the complex plane.  As has been discussed
extensively in the literature \cite{buni}, the unitarity triangle
provides a sensitive test of the standard model, and in particular, of
the CKM picture of CP violation.  If future experiments at B factories
reveal that the unitarity triangle fails to close, it would imply the
existence of new physics.  What has provoked much excitement is the
realization that CP asymmetries in certain exclusive B-meson decays
can be related to the angles of the unitarity triangle in a simple way,
often without hadronic uncertainties.  The $B$ decays of interest are
ones in which the final state can be reached by both  $B^0$ and
$\overline{B}^0$ decay.  Since a $B^0$ meson created at time
zero becomes a mixture of $B^0$ and $\overline{B}^0$ at time $t$, the
decay amplitudes of each component to the final state $f$ interfere,
giving a time dependent asymmetry:
\beq
a = \frac{\Gamma (B^0(t)\rightarrow f)-
\Gamma (\overline{B}^0(t)\rightarrow f)}
{\Gamma (B^0(t)\rightarrow f)+
\Gamma (\overline{B}^0(t)\rightarrow f)}
= \sin \Delta m \,t \,\,{\rm Im}\left(\frac{p}{q}\rho\right)
\label{eq:cpasymm}
\eeq
where in the standard notation
\begin{eqnarray}
\left(\frac{p}{q}\right)^2 = \frac{M_{12} - \frac{i}{2} \Gamma _{12}}
{M_{12}^* - \frac{i}{2} \Gamma _{12}^*} & \mbox{ and } &
\rho = \frac{A(B^0\rightarrow f)}{ A(
\overline{B}^0\rightarrow f)}
\label{eq:pq}
\end{eqnarray}
where $M_{12}$ and $\Gamma_{12}$ are the off-diagonal mass and widths.
Note that the decay amplitudes in the definition of $\rho$ are those
of pure $B^0$ and $\overline{B}^0$ states.  For the $B^0$ system,
$M_{12} \gg \Gamma _{12}$, so $p/q$ simply tells us the phase of the
off-diagonal mass.  Furthermore, if the decay proceeds through
a single weak interaction diagram, and if $f$ is a CP eigenstate, one
can show that $\rho$ is also purely a phase, allowing us to write
\beq
{\rm Im} \, \frac{p}{q} \rho \equiv \sin 2 \Phi
\label{eq:Phi}
\eeq
In the standard model, the angle $\Phi$ can be expressed in terms of
the phases of the CKM matrix elements relevant to the mixing and to the
specific decay channel.  For the decays
$B_S^0\rightarrow \rho K_S$, $B^0 \rightarrow \psi K^0_S$,
and $B^0 \rightarrow \pi^+\pi^-$, it has been shown that $\Phi$ is
identically $\alpha$, $\beta$, and $\gamma$, respectively, the three angles
of the unitarity triangle, as labelled in Fig.\ref{fig:untri1}. \cite{gilman}.

The new four quark interactions in our composite technicolor model
are approximately superweak; they are suppressed by at least a factor
of $G_F/f_2^2$ in comparison to the standard model operators.  Therefore,
we do not expect large corrections to any weak decay amplitudes, except
in the case of rare decays that are induced at one-loop in the standard
model.  For the tree-level decays that are of interest to us here, it
follows that $\rho$ will not be altered from its standard
model value.  On the other hand, $p/q$ tells us the overall phase of the
operators that contribute to the off-diagonal mass $M_{12}$.  If the new
$\Delta B = 2$ operators in our model had a phase differing from the
standard model box diagram, then $p/q$ would be altered, and $\Phi$ would
no longer give us the correct angles of the unitarity
triangle.  However, the operators in the CTSM model that
contribute to $M_{12}$  have the same phase as the top quark box diagram,
and it just happens that this diagram dominates over the other standard
model contributions, as we saw in (\ref{eq:smbd}).  As a result, $p/q$ will be
the same as in the standard model, and the relation between CP asymmetries
and the angles in the unitarity triangle will not be effected.

\begin{figure}
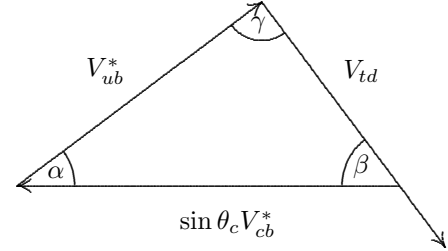

$$
\beginpicture
\setcoordinatesystem units < 0.02in, 0.02in > point at 0 0
\arrow <5pt> [0.3,1] from 100 0 to 0 0
\arrow <5pt> [0.3,1] from 0 0 to 64 48
\arrow <5pt> [0.3,1] from 64 48 to 112 -16
\put{$\alpha$}  at 10 3.5
\put{$\beta$}  at 90  5
\put{$\gamma$}  at 63 42
\circulararc 36.86989 degrees from 15 0 center at 0 0
\circulararc 90.0 degrees from 56 42 center at 64 48
\circulararc -53.13011 degrees from 85 0 center at 100 0
\put{$\sin \theta_c V^*_{cb}$} at 55 -10
\put{$V_{td}$} at 90 30
\put{$V_{ub}^*$} at 23 30
\endpicture
$$
\label{fig:untri2}
\caption{Possible unitarity triangle in the CTSM model.}
\end{figure}

Also notice that two of the sides of the unitarity triangle are
unchanged by the superweak interactions in our model.  Both $|V_{ub}|$
and $|V_{cb}|$ will be determined accurately from semileptonic $B$ decays,
fixing the length of two of the sides. Since two sides and three angles of
the unitarity triangle would be immune to CTSM effects, we would conclude
that the third side has the correct orientation to close the triangle.
However, the length of the third side, $|V_{td}|$, is likely to be
extracted from $B^0$-$\overline{B}^0$ mixing, which is drastically altered
by CTSM effects, as we saw in (\ref{eq:bdfull}).  A unitarity triangle
that is plausible in every way, except for the length of the
side $V_{td}$, as in Fig.\ref{fig:untri2}, would give us indication of the
additional $\Delta B =2$ interactions in the CTSM model.

\section{D-Meson Physics}\label{sec:dphys}

While the CTSM operators can have a potentially large effect
on $B^0$-$\overline{B}^0$ mixing, they do not leave us with a
unique signature for the model.   In the $D^0$ system, however, the
situation may be more promising if the composite technicolor
operators are large.  From (\ref{eq:opup}), the relevant four-quark
interaction is
\beq
\frac{1}{f_1^2} (V_{ub}^* V_{cb})^2\, \frac{ (1- \cos (\phi - \phi '))^2}
{\sin ^4 \phi} \,(\overline{c}_L \gamma^\mu u_L)^2
\label{eq:cunew}
\eeq
Note that there is no contribution to the $\Delta C =2$ operator
from symmetry breaking terms of the form (\ref{eq:mumu}), from
the scale $f_2$.  To get an idea of the size of (\ref{eq:cunew}), we
will estimate its contribution to the $D^0$-$\overline{D}^0$ mass splitting.
Using the characteristic values of the CTSM parameters from Section
\ref{sec:bphys}, and the vacuum insertion approximation to estimate
the matrix element, we find
\beq
\Delta M _{CTSM} \approx  4\times10^{-12} \mbox{ MeV}
\label{eq:dmmm}
\eeq
As we will see below, this result is interesting in that it is potentially
larger than the standard model estimate.   Note that it is well
below the current experimental limit \cite{PDB}
\beq
\Delta M _{exp} < 1.3 \times 10^{-10} \mbox{ MeV}
\eeq

The precise standard model contribution to $\Delta M$, on the
other hand, is far more difficult
to determine.  A number of authors have estimated that the
dispersive contribution to $M_{12}$ in the $D^0$ system from two pion
exchange is more than an order of magnitude larger than the short distance
contribution from the standard model box diagram \cite{donoghue}.
These authors have argued that
\beq
\Delta M _{SM} \approx 10^{-12} \mbox{ MeV}
\label{eq:dmsm}
\eeq
from an estimate of the long distance effects.  If we take this at face value,
we conclude that the $D^0$-$\overline{D}^0$ mass splitting is enhanced in
the CTSM model, though perhaps by not enough for us to distinguish the new
short distance effect from the uncertainty in (\ref{eq:dmsm}).  If this is the
case, then there is not much more we can say.  However, there is a great deal
of theoretical uncertainty in both (\ref{eq:dmmm}) and (\ref{eq:dmsm}).
Our estimate could easily be made an order of magnitude larger by adjusting
the magnitudes of the CKM factors within their allowed ranges, or by
adopting a smaller value for the unknown scale $f_1$.  Furthermore, a recent
reformulation \cite{ddhq} of the standard model estimate in the language
of the heavy quark effective theory has suggested that the results of
ref. \cite{donoghue} may prove to be an overestimate.
With these uncertainties in mind, it is not unreasonable to consider
the possibility that $\Delta M_{CTSM} \approx 10 \,\Delta M_{SM}$.
Assuming that this is the case,  we will ignore the standard model
contribution to $D^0$-$\overline{D}^0$  mixing all together,
and derive our predictions from (\ref{eq:cunew}).

If the CTSM operator is large, then it is the primary source of
indirect CP-violation in the $D^0$ system.   In this limit, we can
carry over the formalism introduced in Section \ref{sec:bphys} to
study the CP violating asymmetries in $B$ decays.  Recall that the
combination of parameters effected by the additional superweak
interactions in the model was $p/q$
\beq
\left(\frac{p}{q}\right)^2 = \frac{M_{12} - \frac{i}{2} \Gamma _{12}}
{M_{12}^* - \frac{i}{2} \Gamma _{12}^*} \approx \left( \frac{M_{12}}
{M_{12}^*} \right)_{CTSM}
\label{eq:dparam}
\eeq
where $M_{12}$ is now the off-diagonal mass for the $D^0$ system.
Notice that the approximate equality in (\ref{eq:dparam}) follows from
our assumption that the new short distance contribution $M_{12}^{CTSM}$
dominates the total standard model contributions to both $M_{12}$ and
$\Gamma_{12}$.  From (\ref{eq:cunew}), we conclude that
\beq
\frac{p}{q} \approx \frac{ V_{ub}^* V_{cb} } {V_{ub} V_{cb}^*}
\eeq
The parameter $\rho$, which is uneffected by the CTSM interactions,
again is most easily evaluated in decays to a CP eigenstate that proceed
predominantly through a single weak interaction amplitude.  For the decay
$D^0 \rightarrow K_S^0 \pi^0$, for example, we obtain
\beq
\rho \approx \frac{V_{us} V_{cd}^*}{V_{us}^* V_{cd}}
\eeq
The CP violating decay asymmetry is then determined by the angle
$\Phi$ defined in (\ref{eq:Phi}),
\beq
\Phi \approx \, \mbox{Arg } \left( V_{ub}^* V_{cb} V_{us} V_{cd}^*
\right)\approx \alpha
\label{eq:dresult}
\eeq
where $\alpha$ is the angle in the unitarity triangle shown
in Fig. \ref{fig:untri1}.  From this example, it is clear that
we would obtain the same result for any appropriately chosen
$D^0$ decay mode;  in the parameterization in which the phase
of $V_{ub}$ is large, the only other CKM factors that enter the
definition of $\Phi$ (from the upper left two by two block
of the CKM matrix) will have phases that are much smaller, and we again
would find that the asymmetry depends on the angle $\alpha$.  With
$\alpha$ determined independently from the CP asymmetry in
$B_s^0 \rightarrow \rho K_S^0$, as discussed in Section \ref{sec:bphys},
our result (\ref{eq:dresult}) provides us with a potentially testable
prediction of the effects of the CTSM operator
on $D^0$ decay asymmetries.  However, as we shall see in the
next section, these decay asymmetries will be quite small, and
difficult to measure.

\section{K-Meson Physics}

As in the $B^0$ system, three operators contribute to the low-energy
$\Delta S = 2$ effective Hamiltonian.  From the symmetry breaking
term (\ref{eq:mumu}) we obtain the operator
\beq
{\cal O}_1^{\Delta S =2} = \pm {G_F^3 \over (4\pi)^4}
		(\xi_t^K m_t^2 + \xi_c^K m_c^2)^2 (f_2 / v)^6
                (\overline{d}_L \gamma^{\mu}s_L)^2
\label{eq:rone}
\eeq
where
\beq
	\xi_q^K \equiv V_{qd}V_{qs}^*.
\eeq
The $\Delta S = 2$ piece of the new operator (\ref{eq:opdown}) is
given by
\beq
{\cal O}_2^{\Delta S =2} =
\frac{1}{f_1^2} (V_{td}^* V_{ts})^2 \frac{ (1-\cos (\phi -\phi ') )^2}
{\sin ^4 \phi} (\overline{d}_L \gamma^{\mu}s_L)^2
\label{eq:rtwo}
\eeq
Finally, we include the contribution from the standard
model box diagram
\beq
{\cal O}_{S.M.}^{\Delta S =2} =
	{G_F^2 \over 16 \pi^2}( (\xi_t^K)^2 m_t^2
		+\xi_t^K \xi_c^K m_c^2 {\rm ln} ( {m_c^2 \over m_t^2} )
		+ (\xi_c^K)^2 m_c^2 ) \, (\overline{d}_L \gamma^{\mu}s_L)^2
\label{eq:rthree}
\eeq
To determine the $K^0$-$\overline{K}^0$ mass splitting, we
adopt the values of the parameters presented in Section \ref{sec:bphys},
and estimate the hadronic matrix element using the vacuum insertion
approximation.  We find
\beq
\Delta M _{CTSM} =  ( \pm 0.6 \, +  3.2 \, + \, 0.6) \times 10^{-15}
\mbox{GeV}
\eeq
where the three terms shown come from (\ref{eq:rone}), (\ref{eq:rtwo}),
and (\ref{eq:rthree}), respectively. Notice that this is consistent
with the experimental result
\beq
\Delta M_{exp} = 3.522 \times 10^{-15} \mbox{GeV}
\eeq
considering the large hadronic uncertainties involved.

The imaginary piece of the $\Delta S = 2$  Hamiltonian contributes to
indirect CP violation in the $K^0$ system, and hence to the
$\epsilon$ parameter.  In the standard phase convention, $\epsilon$
is given by
\beq
\epsilon \approx \frac{e^{i\pi/4} {\rm Im} \, M_{12} }{\sqrt{2} \Delta M}
\label{eq:epdef}
\eeq
It is straightforward to compute the imaginary parts of the
the operators (\ref{eq:rone}), (\ref{eq:rtwo}), and (\ref{eq:rthree}),
which contribute to Im $M_{12}$.  We find
\beq
\left[ \pm 3\times 10^{-7} + 3 \times 10^{-6} + 9 \times 10^{-9} \right]
(\mbox{GeV}^{-2}) \, \mbox{Im}(\xi^K_t)^2 (\overline{s}_L
\gamma^{\mu}d_L)^2
\label{eq:impart}
\eeq
Clearly, the new operator (\ref{eq:rtwo}) dominates (\ref{eq:impart})
and is two orders of magnitude larger than the standard model
contribution.  This will force us to place a tight constraint on the
size of the CP violating phase in the CKM matrix, given the experimental
information on the real part of $\epsilon$.  From (\ref{eq:impart}), we find
\beq
\mbox{Im}\, M_{12} \approx (1\times 10^{-15}) \sin (2 \,
\, \mbox{Arg} \, V_{td})
\eeq
where we have assumed that the phase of $M_{12}$ comes purely from the CKM
factors in the $\Delta S =2$ operator.  Then, we obtain
\beq
\mbox{Re} \, \epsilon \approx (0.15) \sin (2\,\mbox{Arg} \, V_{td})
\eeq
The real part of $\epsilon$ has been extracted from the observed
CP violating charge asymmetry in semileptonic $K_L$ decay \cite{PDB},
\beq
(\mbox{Re}\, \epsilon )_{exp} = 1.635 \times 10^{-3}
\eeq
which leads us to conclude that $(2 \,\mbox{Arg} \, V_{td})$ is of the order
$10^{-2}$.  With CKM phase angles of this size, the time-dependent
decay asymmetries discussed in Sections \ref{sec:bphys} and
\ref{sec:dphys}  should have amplitudes of approximately $1$\%.
While this result is not encouraging from an experimental point of view,
at least we can say that larger asymmetries would immediately
rule out the model.  The smallness of the CKM phase angle also alters
the theoretical prediction for $|\epsilon ' / \epsilon|$.  Standard model
estimates of this ratio that assume a top quark mass of $100$ GeV, and  a
CKM phase of order $1$ give results that are no larger than
${\cal O}(10^{-3})$ \cite{buras};  since our phase is constrained to be two
orders of magnitude smaller, we conclude
\beq
\epsilon ' / \epsilon \leq {\cal O}(10^{-5})
\eeq
While this result is in conflict with the experimental value for
$\epsilon ' / \epsilon$ quoted in the Particle Data Book, we must emphasize
that the experimental results are uncertain.  The global fit to $\epsilon '
/ \epsilon$ quoted by the Particle Data Group is only $2\sigma$ away
from zero \cite{PDB}, with some individual groups claiming measurements
consistent with zero \cite{yam}.  Clearly, future experimental measurements
of $\epsilon ' / \epsilon$ will provide a stringent test of the
viability of this model.

\section{Conclusions}

We have considered flavor changing neutral current effects and CP
violation arising from non-standard four-quark interactions in a
composite technicolor model.  We have shown that the mass
splittings of the CP eigenstates in the $B^0$, $D^0$, and $K^0$
systems may each receive large contributions from the CTSM operators.
While we have discussed in some detail the pattern of small CP violating
decay asymmetries in the $B^0$ and $D^0$ systems due to the composite
technicolor interactions, we have found that the most dramatic implication of
the model is that $\epsilon ' / \epsilon$ in the $K^0$ system should be
two orders of magnitude smaller than most standard model estimates.
We suggest that this result may prove fatal to this model if experiments
conclusively establish a large non-zero value for $\epsilon ' / \epsilon$.

\vspace{36pt}
\centerline{\bf Acknowledgments}
We are grateful to Howard Georgi for suggesting this
phenomenological investigation.  We thank Lisa Randall
for useful comments.
{\em Research supported
in part by the National Science Foundation under Grant
PHY-8714654, and in part by the Texas National Research Laboratory
Commission, under Grant RGFY9206.}


\end{document}